\documentclass[prc,twocolumn,nofootinbib,superscriptaddress,showpacs]{revtex4}
\usepackage{graphicx,amsmath,amssymb,bm,multirow}

\newcommand{\be}{\begin{equation}}
\newcommand{\ee}{\end{equation}}
\newcommand{\bea}{\begin{eqnarray}}
\newcommand{\eea}{\end{eqnarray}}
\newcommand{\ba}{\begin{align}}
\newcommand{\ea}{\end{align}}

\newcommand{\vlowk}{V_{{\rm low}\,k}}
\newcommand{\vbar}{\overline{V}_{\rm 3N}}
\newcommand{\lm}{\Lambda}
\newcommand{\kf}{k_{\rm F}}
\newcommand{\tr}{{\rm Tr}}
\newcommand{\fm}{\, \text{fm}}
\newcommand{\fmi}{\, \text{fm}^{-1}}
\newcommand{\fmiq}{\, \text{fm}^{-3}}
\newcommand{\mev}{\, \text{MeV}}
\newcommand{\kev}{\, \text{keV}}
\newcommand{\gevi}{\, \text{GeV}^{-1}}
\newcommand{\la}{\langle}
\newcommand{\ra}{\rangle}
\newcommand{\tl}{\widetilde{l}}
\newcommand{\tlp}{\widetilde{l'}}
\newcommand{\tj}{\widetilde{J}}

\begin{document}

\title{Chiral three-nucleon forces and neutron matter}

\author{K.\ Hebeler}
\email[E-mail:~]{hebeler@triumf.ca}
\affiliation{TRIUMF, 4004 Wesbrook Mall, Vancouver, BC, V6T 2A3, Canada}
\author{A.\ Schwenk}
\email[E-mail:~]{schwenk@triumf.ca}
\affiliation{TRIUMF, 4004 Wesbrook Mall, Vancouver, BC, V6T 2A3, Canada}
\affiliation{ExtreMe Matter Institute EMMI, GSI Helmholtzzentrum f\"ur
Schwerionenforschung GmbH, 
64291 Darmstadt, Germany}
\affiliation{Institut f\"ur Kernphysik, Technische Universit\"at
Darmstadt, 
64289 Darmstadt, Germany}

\begin{abstract}
We calculate the properties of neutron matter and highlight the
physics of chiral three-nucleon forces. For neutrons, only the
long-range $2 \pi$-exchange interactions of the leading chiral
three-nucleon forces contribute, and we derive density-dependent
two-body interactions by summing the third particle over occupied
states in the Fermi sea. Our results for the energy suggest that
neutron matter is perturbative at nuclear densities. We study
in detail the theoretical uncertainties of the neutron matter
energy, provide constraints for the symmetry energy and its 
density dependence, and explore the impact of chiral three-nucleon
forces on the $S$-wave superfluid pairing gap.
\end{abstract}


\maketitle

\section{Introduction}

The physics of neutron matter ranges over exciting extremes: from
universal properties at low densities~\cite{dEFT,Gezerlis} that can be
probed in experiments with ultracold atoms~\cite{coldatoms}; to using
neutron matter properties at nuclear densities to guide the
development of a universal density functional~\cite{DFT1,DFT2} and to
constrain the physics of neutron-rich nuclei; to higher densities
involved in the structure of neutron stars~\cite{LP}. In the theory of
nuclear matter, recent advances~\cite{nucmatt,chiralnm} are based on
systematic chiral effective field theory (EFT)
interactions~\cite{chiral,RMP} combined with a renormalization group
(RG) evolution to low momenta~\cite{Vlowk,smooth}. This evolution
improves the convergence of many-body calculations~\cite{nucmatt,%
NCSM,Sonia} and the nuclear matter energy shows saturation
with controlled uncertainties~\cite{chiralnm}. In this paper, we
extend theses developments to neutron matter with a focus on
three-nucleon (3N) forces.

Our studies are based on evolved nucleon-nucleon (NN) interactions
at next-to-next-to-next-to-leading order (N$^3$LO)~\cite{N3LO,N3LOEGM}
and on the leading N$^2$LO 3N forces~\cite{chiral3N1,chiral3N2}. In
Sect.~\ref{densdep}, we show that only the long-range $2\pi$-exchange
3N interactions contribute in pure neutron matter. We then construct
density-dependent two-body interactions $\vbar$ by summing the third
particle over occupied states in the Fermi sea. Effective interactions
of this sort have been studied in the past using 3N potential
models and approximate treatments (see for example,
Refs.~\cite{3Nold1,3Nold2}). We derive a general operator and momentum
structure of $\vbar$ and analyze the partial-wave contributions and
the density dependence of $\vbar$. This provides insights to the role
of chiral 3N forces in neutron matter.

In Sect.~\ref{results}, we apply $\vbar$ to calculate the properties
of neutron matter as a function of Fermi momentum $\kf$ (or the
density $\rho = \kf^3/(3\pi^2)$) based on a loop expansion around the
Hartree-Fock energy. Our second-order results for the energy suggest
that neutron matter is perturbative at nuclear densities, where
N$^2$LO 3N forces provide a repulsive contribution. We study in detail
the theoretical uncertainties of the neutron matter energy and find
that the uncertainty in the $c_3$ coefficient of 3N forces
dominates. Other recent neutron matter calculations lie within the
resulting energy band. In addition, the energy band provides
constraints for the symmetry energy and its density
dependence. Finally, we study the impact of chiral 3N forces on the
$^1$S$_0$ superfluid pairing gap at the BCS level. We conclude and
give an outlook in Sect.~\ref{concl}.

\section{Three-nucleon forces as density-dependent two-body interactions}
\label{densdep}

Nuclear forces depend on a resolution scale, which is generally
determined by a momentum cutoff $\lm$, and are given by an effective
theory for scale-dependent two- and corresponding many-nucleon
interactions~\cite{chiral,RMP,Vlowk,pionless}:
\be
H(\lm) = T + V_{\rm NN}(\lm) + V_{\rm 3N}(\lm) + V_{\rm 4N}(\lm)
+ \ldots \,.
\ee
Our calculations are based on chiral EFT
interactions. We start from the N$^3$LO NN potential ($\lm=500 \mev$)
of Ref.~\cite{N3LO} and use the RG to evolve
this NN potential to low-momentum interactions $\vlowk$ with a smooth
$n_{\rm exp}=4$ regulator with $\lm = 1.8-2.8
\fmi$~\cite{smooth,Kai}. This evolution softens the short-range
repulsion and short-range tensor components of the initial
chiral interaction~\cite{nucmatt,Born}. Based on the universality of
$\vlowk$~\cite{chiralnm,smooth}, we do not expect large differences
starting from different N$^3$LO potentials.

In chiral EFT without explicit Deltas, 3N forces start at N$^2$LO
and contain a long-range $2 \pi$-exchange part $V_c$, an
intermediate-range $1 \pi$-exchange part $V_D$ and a short-range
contact interaction $V_E$~\cite{chiral3N1,chiral3N2}:
\be
\parbox[c]{195pt}{%
\includegraphics[scale=0.6,clip=]{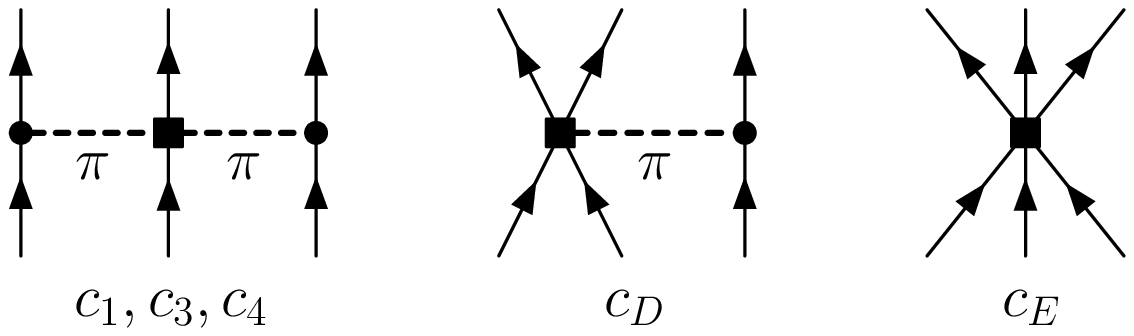}}
\label{3NF}
\ee
The $2 \pi$-exchange interaction is given by
\be
V_c = \frac{1}{2} \, \biggl( \frac{g_A}{2 f_\pi} \biggr)^2 
\sum\limits_{i \neq j \neq k} 
\frac{({\bm \sigma}_i \cdot {\bf q}_i) ({\bm \sigma}_j \cdot 
{\bf q}_j)}{(q_i^2 + m_\pi^2) (q_j^2 + m_\pi^2)} \: F_{ijk}^{\alpha\beta} \,
\tau_i^\alpha \, \tau_j^\beta \,,
\label{Vc}
\ee
where ${\bf q}_i = {\bf k}'_i - {\bf k}_i$ denotes the difference of initial
and final nucleon momenta ($i, j$ and $k=1,2,3$) and
\begin{multline}
F_{ijk}^{\alpha\beta} = \delta^{\alpha\beta} \biggl[ - \frac{4 c_1 
m_\pi^2}{f_\pi^2} + \frac{2 c_3}{f_\pi^2} \: {\bf q}_i \cdot {\bf q}_j
\biggr] \\[1mm]
+ \sum_\gamma \, \frac{c_4}{f_\pi^2} \: \epsilon^{\alpha\beta\gamma}
\: \tau_k^\gamma \: {\bm \sigma}_k \cdot ( {\bf q}_i \times {\bf q}_j)
\,, \label{Vc2}
\end{multline}
while the $1 \pi$-exchange and contact interactions are given respectively by
\begin{align}
V_D &= - \frac{g_A}{8 f_\pi^2} \, \frac{c_D}{f_\pi^2 \lm_\chi}
\: \sum\limits_{i \neq j \neq k} 
\frac{{\bm \sigma}_j \cdot {\bf q}_j}{q_j^2 + m_\pi^2} \: ({\bm \tau}_i
\cdot {\bm \tau}_j) \, ({\bm \sigma}_i \cdot {\bf q}_j) \,, 
\label{VD} \\
V_E &= \frac{c_E}{2 f_\pi^4 \lm_\chi} \: \sum\limits_{j \neq k} ({\bm \tau}_j
\cdot {\bm \tau}_k) \,, \label{VE}
\end{align}
with $g_A = 1.29$, $f_\pi = 92.4 \mev$, $m_\pi = 138.04 \mev$ and
$\lm_\chi = 700 \mev$. For 3N interactions, we use a smooth regulator
as in Ref.~\cite{chiralnm},
\be
f_{\text{R}}(p,q) = \exp \biggl[ - \frac{(p^2+3 q^2/4)^2}{\lm_{\rm 3NF}^4}
\biggr] \,,
\label{reg}
\ee
with a 3N cutoff $\lm_{\rm 3NF}$ that is allowed to vary
independently of the NN cutoff and probes short-range three-body
physics. Here, $p$ and $q$ are initial Jacobi
momenta (the final Jacobi momenta are denoted by $p'$ and $q'$).
The exchange terms of the 3N force are included by means of the
antisymmetrizer
\begin{align}
{\mathcal A}_{123} &= (1 + P_{12} P_{23} + P_{13} P_{23}) (1 - P_{23}) 
\,, \label{antisym1} \\[1mm]
&= 1 - P_{12} - P_{13} - P_{23} + P_{12} P_{23} + P_{13} P_{23} \,,
\label{antisym2}
\end{align}
where $P_{ij}$ is the exchange operator for spin, isospin and momenta 
of nucleons $i$ and $j$.  The regulator $f_{\text{R}}(p,q)$ is totally
symmetric when expressed in the nucleon momenta ${\bf k}_i$, and thus
the direct and exchange terms contain the same regulator.

In this work, we take the N$^2$LO 3N forces as a truncated basis for
low-momentum 3N interactions and assume that the $c_i$ coefficients of
the long-range $2 \pi$-exchange part $V_c$ are not modified by the RG
evolution. This follows the strategy adapted in
Refs.~\cite{nucmatt,chiralnm,Vlowk3N}, until it will be possible to
evolve many-body forces in momentum space starting from chiral EFT.
For chiral low-momentum interactions, the $c_D$ and $c_E$ couplings
have been fit for various cutoffs to the $^3$H binding energy and the
$^4$He radius in Ref.~\cite{chiralnm}. However, as will be shown in
the following, in pure neutron matter only the $c_1$ and $c_3$ terms
of the $2 \pi$-exchange part $V_c$ contribute, so that the leading
low-momentum three-neutron interactions will not depend on the
shorter-range parts. Since the $c_D$ and $c_E$ couplings do not enter
in neutron matter, we take for the $c_i$ coefficients the consistent
values used in the N$^3$LO NN potential of Ref.~\cite{N3LO}, in
particular $c_1 = -0.81 \gevi$ and $c_3 = -3.2 \gevi$.

In addition, we will estimate the uncertainties of the two assumptions
we have made for 3N forces. First, we will vary the cutoff to probe
the sensitivity to neglected short-range many-body interactions and
to the completeness of the many-body calculation. Second, we will study
the dependence of our results on the choice of the $c_i$ coefficients
within their theoretical uncertainties in Sect.~\ref{ciuncertain}.

\subsection{Chiral 3N forces in neutron matter}
\label{3Nneutmatt}

\begin{figure*}[t]
\parbox{1.2cm}{\centering \includegraphics[scale=0.8]{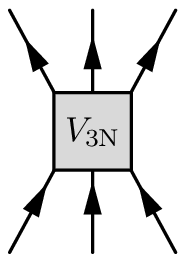}} 
\parbox{0.6cm}{\centering{$=$}} 
\parbox{2.0cm}{\centering \includegraphics[scale=0.6]{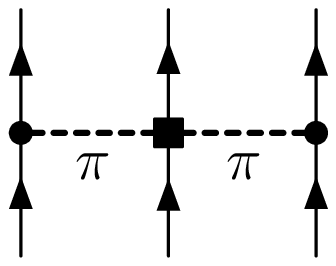}} 
\parbox{0.4cm}{\centering{$-$}} 
\parbox{2.0cm}{\centering \includegraphics[scale=0.6]{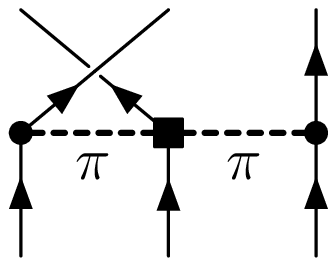}} 
\parbox{0.4cm}{\centering{$-$}} 
\parbox{2.0cm}{\centering \includegraphics[scale=0.6]{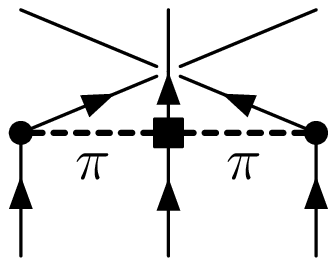}} 
\parbox{0.4cm}{\centering{$-$}}
\parbox{2.0cm}{\centering \includegraphics[scale=0.6]{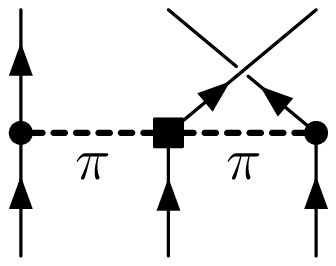}}
\parbox{0.4cm}{\centering{$+$}} 
\parbox{2.0cm}{\centering \includegraphics[scale=0.6]{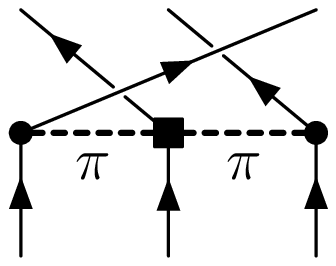}}
\parbox{0.4cm}{\centering{$+$}}
\parbox{2.0cm}{\centering \includegraphics[scale=0.6]{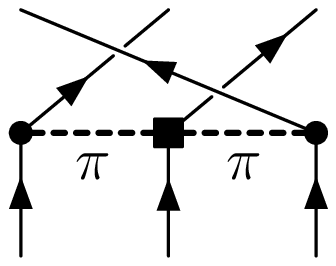}}
\vspace*{-2mm}
\caption{Diagrammatic representation of the antisymmetrized N$^2$LO
3N force in neutron matter. The $c_4$ term in $V_c$ and the
shorter-range parts $V_D$ and $V_E$ vanish in neutron matter.
\label{fig:3Nneut}}
\end{figure*}

Neutron matter presents a very interesting system, because only
certain parts of the N$^2$LO 3N forces contribute (see, for instance,
Ref.~\cite{neutmatt}). First, the contact interaction $V_E$ vanishes
between antisymmetrized states, because three neutrons cannot interact
in relative $S$-wave states due to the Pauli principle.

Second, the antisymmetrized $1 \pi$-exchange part $V_D$ vanishes due
to the particular spin-momentum structure of this interaction. With
${\bm \tau}_i \cdot {\bm \tau}_j = 1$ for neutrons, one has
\be
\mathcal{A}_{123} V_D \bigr|_{nnn}
\sim \mathcal{A}_{123} \sum\limits_{i \neq j \neq k} 
\frac{{\bm \sigma}_j \cdot {\bf q}_j \, {\bm \sigma}_i \cdot {\bf
q}_j}{q_j^2 + m_\pi^2} \,,
\label{VDneut}
\ee
where in Eq.~(\ref{VDneut}) and in the following the isospin exchange
operator $P_{ij}^\tau = 1$ in the antisymmetrizer.
The sum over $i \neq j \neq k$ can be grouped into three terms, ${\bm
\sigma}_1 \cdot {\bf q}_1 \, ({\bm \sigma}_2 + {\bm \sigma}_3) \cdot
{\bf q}_1$, and with $1 \, (23)$ replaced by $2 \, (13)$ and $3 \,
(12)$.  The first of these three terms is independent of the momenta
of particles $2$ and $3$ and is nonvanishing only if the spin part of
the wave function is symmetric under the exchange of the spins of
particles $2$ and $3$. This implies $P_{23} = 1$ and using the product
representation of the antisymmetrizer, Eq.~(\ref{antisym1}), leads to
\be
\mathcal{A}_{123} \: {\bm \sigma}_1 \cdot {\bf q}_1 \, ({\bm \sigma}_2
+ {\bm \sigma}_3) \cdot {\bf q}_1 = 0 \,.
\ee
The other two terms in $V_D$ vanish similarly. The physical reason
for $\mathcal{A}_{123} V_D \bigr|_{nnn} =0$ is that the two 
particles interacting through the contact interaction $c_D$ of
Eq.~(\ref{3NF}) are required to be in a symmetric two-body spin 
state due the structure ${\bm \sigma}_i + {\bm \sigma}_j$ (because
pion exchange couples to the sum of the spins), but the
interaction does not depend on the momenta of these two particles,
which means the momentum-space wave function will be symmetric. Therefore,
the two particles cannot be in an overall antisymmetric state.

Finally, the $c_4$ term in Eq.~(\ref{Vc2}) of the $2 \pi$-exchange
part $V_c$ does not contribute for neutrons, because of the isospin
structure with $\la nnn | {\bm \tau}_1 \cdot ({\bm \tau}_2 \times {\bm
\tau}_3) | nnn \ra = 0$. As a result, the N$^2$LO 3N force in neutron
matter is given by
\begin{multline}
\mathcal{A}_{123} V_{\rm 3N} \bigl|_{nnn}
= f_{\text{R}}(p',q') f_{\text{R}}(p,q) \,
\frac{g_A^2}{4 f_\pi^4} \, \mathcal{A}_{123} \\[1mm]
\times \sum\limits_{i \neq j \neq k} 
\frac{({\bm \sigma}_i \cdot {\bf q}_i) ({\bm \sigma}_j \cdot 
{\bf q}_j)}{(q_i^2 + m_\pi^2) (q_j^2 + m_\pi^2)} 
\bigl[ - 2 c_1 m_\pi^2 + c_3 \, {\bf q}_i \cdot {\bf q}_j \bigr] \,.
\label{3Nneut}
\end{multline}
The direct, single- and double-exchange terms included in
Eq.~(\ref{3Nneut}) are shown diagrammatically in
Fig.~\ref{fig:3Nneut}. Since $c_3$ is typically a factor $4$ larger
than $c_1$ and because ${\bf q}_i \cdot {\bf q}_j$ increases with
density compared to $m_\pi^2$ (for densities of interest $q \sim \kf
\sim 2 m_\pi$), the N$^2$LO 3N force in neutron matter is therefore
dominated by the $c_3$ contribution.

\subsection{Operator structure of $\vbar$}
\label{operator}

Based on the N$^2$LO 3N force, Eq.~(\ref{3Nneut}), we construct
an antisymmetrized density-dependent two-body interaction $\vbar$ in
neutron matter by summing the third particle over occupied states in
the Fermi sea,
\be
\vbar= \tr_{\sigma_3} \int \frac{d{\bf k}_3}{(2\pi)^3}
\: n_{{\bf k}_3} \, \mathcal{A}_{123} V_{\rm 3N}  \bigl|_{nnn} \,,
\label{vbar}
\ee
where the trace is over the spin of the third particle and $n_{\bf k}$
denotes the Fermi-Dirac distribution function at zero temperature,
$n_{\bf k} = \theta(\kf - |{\bf k}|)$. The antisymmetrization in
Eq.~(\ref{vbar}) can also be written in a more symmetric form using
normalized and antisymmetrized three-body states, $|123\rangle_{\rm as}
= \frac{1}{\sqrt{6}} \, \mathcal{A}_{123} |123\rangle$, so that
$\langle 1'2'3'|\mathcal{A}_{123} V_{\rm 3N}|123\rangle = {_{\rm as}\langle 
1'2'3'|V_{\rm 3N}|123\rangle}_{\rm as}\,$. Diagrammatically, Eq.~(\ref{vbar})
corresponds to connecting one incoming and one outgoing line in the
antisymmetrized 3N force with a noninteracting propagator:
\be
\parbox[c]{1.8cm}{\centering \includegraphics[scale=0.8]{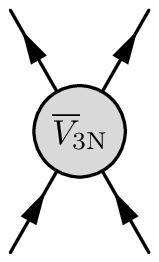}}
\parbox{-0.2cm}{\centering{$=$}} 
\parbox[c]{1.8cm}{\centering \includegraphics[scale=0.8]{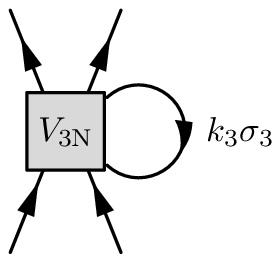}}
\label{vbardiag}
\ee
The resulting density-dependent two-body interaction $\vbar$
corresponds to the normal-ordered two-body part of 3N forces (for
normal-ordering, see Refs.~\cite{Negele,nord}), where the
normal ordering is with respect to the Fermi sea. As a result, the
antisymmetrized $\vbar$ is added to the antisymmetrized NN interaction
$V_{\rm NN,as} = (1-P_{12}) \, V_{\rm NN}$ when evaluating two-body
contributions beyond the Hartree-Fock level. For the Hartree-Fock (or
normal-ordered zero-body) contribution, the 3N diagram has a symmetry
factor $1/6$, so that $\vbar/3$ is added to $V_{\rm NN,as}$ at the
Hartree-Fock level (see Sect.~\ref{HF}). For one-body
contributions, normal ordering leads to $\vbar/2$ being added to
$V_{\rm NN,as}$, for example, for the first-order contribution to the
self-energy. These factors ensure the correct symmetry factors for
diagrams involving NN and 3N interactions. The above normal-ordering
factors may not always be taken into account properly in the
literature~\cite{lit1,lit2,lit3} (for example, it is incorrect to use
$V_{\rm NN,as} + \vbar/3$ or $V_{\rm NN,as} + \vbar$ for both
Hartree-Fock and higher-order contributions).

In addition to the density dependence, the two-body interaction
$\vbar$ depends on the incoming and outgoing relative momenta,
${\bf k} = ({\bf k}_1-{\bf k}_2)/2$ and ${\bf k}' = ({\bf k}'_1-
{\bf k}'_2)/2$, on the spin of particles $1$ and $2$, and on
the two-body center-of-mass momentum ${\bf P}={\bf k}_1+
{\bf k}_2={\bf k}'_1+{\bf k}'_2$. However, as will be shown in
Sect.~\ref{HF}, the $P$ dependence is weak compared to the
dependence on the relative momenta. This can be understood from
the momentum transfers in the pion propagators of 
Fig.~\ref{fig:3Nneut}. The momentum transfers in the first
two diagrams are ${\bf k}-{\bf k}'$, $0$ (for the left
and right propagators) and ${\bf k}+{\bf k}'$, $0$. While these
terms vanish in the considered case of $V_c$, the only
weak $P$ dependence would arise from the regulator functions
in Eq.~(\ref{3Nneut}). For the remaining four diagrams in
Fig.~\ref{fig:3Nneut}, the momentum transfers are ${\bf k} \pm
{\bf k}'$, ${\bf P}/2 \pm {\bf k}' - {\bf k}_3$ and 
${\bf P}/2 + {\bf k} - {\bf k}_3$, ${\bf P}/2 \pm {\bf k}' 
- {\bf k}_3$. Therefore, the $P$ dependence is only through
${\bf k}'_3 = {\bf P}/2 - {\bf k}_3$ and can be transformed
into a dependence due to $n_{{\bf P}/2 - {\bf k}'_3}$  (by
substitution in the integral) and through the regulator functions.
For typical center-of-mass momenta $P \lesssim \kf$, this
results in the weak $P$ dependence.

For the derivation of $\vbar$, we use Mathematica and automate the
momentum and spin exchange operations by representing all spin
operators in matrix form. The antisymmetrized 3N force in neutron
matter, Eq.~(\ref{3Nneut}), is represented in this explicit
three-particle spin basis and for general particle momenta ${\bf
k}_i$. We then perform the spin trace over the third particle and
project the resulting two-body matrix on a complete set of two-body
spin operators.  In this way, we obtain a general operator structure
of $\vbar$ in spin-saturated neutron matter in terms of a set of basic
integral functions. Only at this point we transform $\vbar$ to the
relative and center-of-mass basis and adapt a fixed-$P$
approximation. Since the number of spin operators is minimal for
$P=0$, we take $P=0$ for simplicity and our final result is given by
\be
\vbar = \frac{g^2_A}{4 f^4_\pi} \biggl[ - 2 c_1 m_\pi^2 \,
A_{12}({\bf k},{\bf k}') + c_3 \, B_{12}({\bf k},{\bf k}') \biggr] \,,
\label{3Nnm}
\ee
where the functions $A_{12}({\bf k},{\bf k}')$ and $B_{12}({\bf k},
{\bf k}')$ include all spin dependences,
\begin{widetext}
\begin{align}
A_{12}({\bf k},{\bf k}') &= 2 \, \Bigl[ \rho_+^2({\bf k},{\bf k}')
+ 2 a^1({\bf k},{\bf k}') - a^1({\bf k},-{\bf k}')
- \overline{b}^1({\bf k},{\bf k}') \Bigr] \nonumber \\[1mm]
&- \frac{2}{3} \, {\bm \sigma}_1 \cdot {\bm \sigma}_2 \,
\Bigr[ 2 \rho_-^2({\bf k},{\bf k}') + \rho_+^2({\bf k},{\bf k}')
+ 3 a^1({\bf k},-{\bf k}') - \overline{b}^1({\bf k},{\bf k}')
- 2 \overline{b}^1({\bf k},-{\bf k}') \Bigr] \nonumber \\[1mm]
&+ 4 \, \Bigl[ S_{12}({\bf k} + {\bf k}') \, \rho_+^0({\bf k},{\bf k}')
- S_{12}({\bf k} - {\bf k}') \, \rho_-^0({\bf k},{\bf k}') \Bigr]
- 4 \, {\bm \sigma}_1^a {\bm \sigma}_2^b \,
\Bigl[ \overline{d}^0_{ab}({\bf k},{\bf k}') 
- \overline{d}^0_{ab}({\bf k},-{\bf k}') \Bigr] \nonumber \\[1mm]
&- 2 i \, ({\bm \sigma}_1 + {\bm \sigma}_2)^a \,
\Bigl[ c^0_a({\bf k},{\bf k}') - c^0_a({\bf k},-{\bf k}') \Bigr]
\,, \label{A12} \\[2mm]
B_{12}({\bf k},{\bf k}') &= - 2 \, \Big[ \rho^4_+({\bf k},{\bf k}')
+ 2 a^2({\bf k},{\bf k}') - a^2({\bf k},-{\bf k}')
- \overline{b}^2({\bf k},{\bf k}') \Big] \nonumber \\[1mm]
&+ \frac{2}{3} \, {\bm \sigma}_1 \cdot {\bm \sigma}_2 \,
\Bigr[ 2 \rho_-^4({\bf k},{\bf k}') + \rho_+^4({\bf k},{\bf k}')
+ 3 a^2({\bf k},-{\bf k}') - \overline{b}^2({\bf k},{\bf k}')
- 2 \overline{b}^2({\bf k},-{\bf k}') \Big] \nonumber \\[1mm]
&- 4 \, \Bigl[ S_{12}({\bf k} + {\bf k}') \, \rho_+^2({\bf k},{\bf k}')
- S_{12}({\bf k} - {\bf k}') \, \rho_-^2 ({\bf k},{\bf k}') \bigr]
+ 4 \, {\bm \sigma}_1^a {\bm \sigma}_2^b \, 
\Big[ \overline{d}^1_{ab}({\bf k},{\bf k}') 
- \overline{d}^1_{ab}({\bf k},-{\bf k}') \Big] \nonumber \\[1mm]
&+ 2 i \, ({\bm \sigma}_1 + {\bm \sigma}_2)^a \,
\Bigl[ c^1_a({\bf k},{\bf k}') - c^1_a({\bf k},-{\bf k}') \Big] \,,
\label{B12}
\end{align}
and the basic integral functions are defined by
\begin{align}
\rho_{\pm}^n({\bf k},{\bf k}') &=
\frac{({\bf k} \pm {\bf k}')^n}{\bigl(({\bf k} \pm {\bf k}')^2
+ m_\pi^2\bigr)^2} \int_{{\bf k}_3} 1 \,, \\[1mm]
a^n({\bf k},{\bf k}') &= \int_{{\bf k}_3}
\frac{\bigl( ({\bf k} + {\bf k}_3) \cdot ({\bf k}' + {\bf k}_3)
\bigr)^n}{\bigl( ({\bf k} + {\bf k}_3)^2 + m_{\pi}^2 \bigr) \bigl(
({\bf k}' + {\bf k}_3)^2 + m_{\pi}^2 \bigr)} \,, \\[1mm]
b^n({\bf k},{\bf k}') &= \int_{{\bf k}_3} 
\frac{\bigl( ({\bf k} + {\bf k}') \cdot ({\bf k} + {\bf k}_3) 
\bigr)^n}{\bigl( ({\bf k} + {\bf k}')^2 + m_{\pi}^2 \bigr) \bigl(
({\bf k} + {\bf k}_3)^2 + m_{\pi}^2 \bigr)} \,, \\[1mm]
c^n_a({\bf k},{\bf k}') &= \int_{{\bf k}_3} 
\frac{\bigl( ({\bf k} + {\bf k}_3) \cdot ({\bf k}' + {\bf k}_3) 
\bigr)^{n} \bigl( ({\bf k} + {\bf k}_3) \times ({\bf k}' + {\bf k}_3)
\bigr)_a}{\bigl( ({\bf k} + {\bf k}_3)^2 + m_{\pi}^2 \bigr)
\bigl( ({\bf k}' + {\bf k}_3)^2 + m_{\pi}^2 \bigr)}
\,, \\[1mm]
d^n_{ab}({\bf k},{\bf k}') &= \int_{{\bf k}_3}
\bigl( ({\bf k} + {\bf k}') \cdot ({\bf k} + {\bf k}_3) \bigr)^{n} \,
\frac{({\bf k} + {\bf k}')_a ({\bf k} + {\bf k}_3)_b 
+ ({\bf k} + {\bf k}')_b ({\bf k} + {\bf k}_3)_a - \frac{2}{3}
\, \delta_{ab} \, ({\bf k} + {\bf k}') \cdot 
({\bf k} + {\bf k}_3)}{2 \, \bigl( ({\bf k} + {\bf k}')^2 + m_{\pi}^2
\bigr) \bigl( ({\bf k} + {\bf k}_3)^2 + m_{\pi}^2 \bigr)} \,.
\label{dfunction}
\end{align}
\end{widetext}
In Eqs.~(\ref{A12})--(\ref{dfunction}), the indices $a, b$ run over
the three components of the spin operators, the tensor operator is
given by $S_{12}({\bf p}) = ({\bm \sigma}_1 \cdot {\bf p}) ({\bm
\sigma}_2 \cdot {\bf p}) - p^2 \, {\bm \sigma}_1 \cdot {\bm
\sigma}_2 /3$, and the overline denotes a symmetrization in
the relative momentum variables, $\overline{x}({\bf k}, {\bf k}') =
x({\bf k}, {\bf k}') + x({\bf k}', {\bf k})$. In addition, we have
introduced the short-hand notation for the integration over the
momentum of the third particle,
\be
\int_{{\bf k}_3} = \int \frac{d{\bf k}_3}{(2\pi)^3} \: n_{{\bf k}_3}
\, \widetilde{f}_{\text{R}}(k',k_3) \, \widetilde{f}_{\text{R}}(k,k_3) \,,
\ee
where the regulator, Eq.~(\ref{reg}), expressed in terms of the
relative and third-particle momenta for $P=0$ is given by
$\widetilde{f}_{\text{R}}(k,k_3) = \exp[- (k^2 + k_3^2/3)^2/ \lm_{\rm
3NF}^4]$.

The density-dependent two-body interaction $\vbar$ includes all spin
structures that are invariant under combined rotations in spin and
space in a spin-saturated system. In addition to the central
spin-independent and spin-spin (${\bm \sigma}_1 \cdot {\bm \sigma}_2$)
interactions, the functions $A_{12}$ and $B_{12}$ include tensor
forces ($S_{12}$), spin-orbit interactions ($c^n_a$-terms) and
additional tensor structures ($d^n_{ab}$-terms), which can be
expressed in terms of $S_{12}$ and quadratic spin-orbit
interactions. Finally, we have checked that $\vbar$ is even (odd) in
the scattering angle $\theta_{{\bf k},{\bf k}'}$ in the two-body spin
$S=0$ ($S=1$) channel.

\subsection{Partial-wave matrix elements of $\vbar$}
\label{pw}

\begin{figure}[t]
\begin{center}
\includegraphics[scale=0.9,clip=]{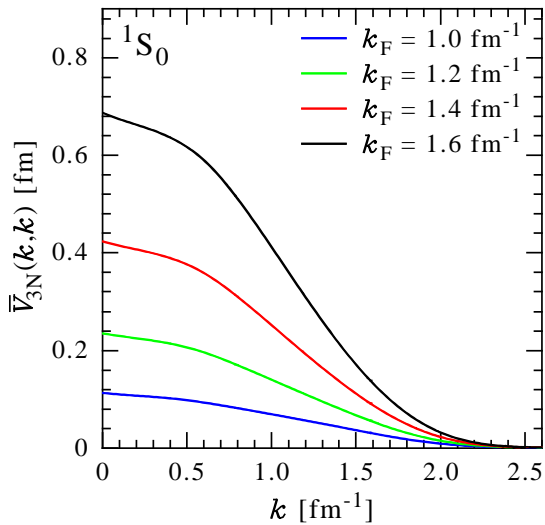}
\end{center}
\caption{(Color online) Diagonal momentum-space matrix elements of 
the density-dependent two-body interaction $\vbar$ for $P=0$ in the
$^1$S$_0$ channel. Results with $\lm_{\rm 3NF} = 2.0 \fmi$ are shown
versus relative momentum $k$ for different Fermi momenta $\kf =
1.0, 1.2, 1.4$ and $1.6 \fmi$ (increasing in strength) in neutron
matter.\label{fig:vbar1S0}}
\end{figure}

\begin{figure}[t]
\begin{center}
\includegraphics[scale=0.9,clip=]{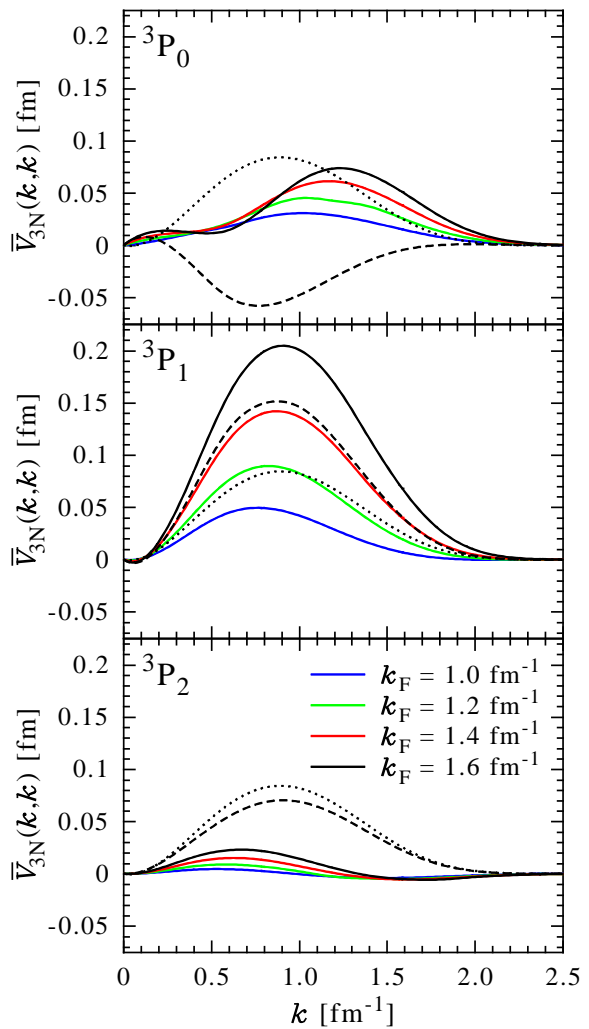}
\end{center}
\caption{(Color online) Diagonal momentum-space matrix elements of 
the density-dependent two-body interaction $\vbar$ for $P=0$ in the
spin-triplet $P$-wave channels. Results with $\lm_{\rm 3NF} = 2.0 \fmi$ are
shown versus relative momentum $k$ for different Fermi momenta
$\kf = 1.0, 1.2, 1.4$ and $1.6 \fmi$ (increasing in strength).
For $\kf = 1.6 \fmi$, the dotted
lines represent the central parts (degenerate in $J$) of $\vbar$, whereas
the dashed lines include the central plus tensor interactions (without the
$c^n_a$-terms in Eqs.~(\ref{A12}) and~(\ref{B12})).\label{fig:vbar3P}}
\end{figure}

Next, we expand $\vbar$ in two-body partial waves and show the
diagonal momentum-space matrix elements in Figs.~\ref{fig:vbar1S0}
and~\ref{fig:vbar3P} in the $^1$S$_0$ and the spin-triplet $P$-wave
channels for different Fermi momenta $\kf = 1.0, 1.2, 1.4$ and $1.6
\fmi$ (in units where $\hbar^2/m = 1$, with
nucleon mass $m$). The partial-wave matrix elements are normalized to
the direct term (as for NN interactions). In Figs.~\ref{fig:vbar1S0}
and~\ref{fig:vbar3P}, we have interpolated between the momentum mesh
points. To set the scale, a typical $S$-wave strength for low-momentum
NN interactions is $\vlowk(0,0) \sim - 2.0 \fm$~\cite{Vlowk,smooth},
but the relative 3N strength depends on whether $\vbar$ is used at the
Hartree-Fock, one- or two-body level (see Sects.~\ref{HF}
to~\ref{2nd}). We find that the dominant partial wave is the
$^1$S$_0$ channel, which is found to be repulsive. In
addition, for $\kf = 1.6 \fmi$ we plot in Fig.~\ref{fig:vbar3P} the
central and central plus tensor parts of $\vbar$. 
Because $S_{12}({\bf p})/p^2 = -4/3, 2/3$ and $-2/15$ in the
$^3$P$_0$, $^3$P$_1$ and $^3$P$_2$ channels, we conclude from
Fig.~\ref{fig:vbar3P} that the tensor force in $\vbar$ is repulsive.
Moreover, with ${\bf L} \cdot {\bf S} = -2, -1$ and $1$ in the
$^3$P$_0$, $^3$P$_1$ and $^3$P$_2$ channels, we observe that the
spin-orbit interaction in $\vbar$ (determined by the difference
of the solid and dashed lines) is attractive.

The density dependence of the two-body matrix elements depends on the
partial wave. In the $^1$S$_0$ channel this dependence can be
approximately parameterized by a power of the Fermi momentum,
$\vbar(k',k,\kf) \approx F(k',k) \, \bigl( \kf/\overline{k}_{\rm F}
\bigr)^4$, with some reference Fermi momentum $\overline{k}_{\rm F}$.
In the spin-triplet channels this is more complex due to the different
momentum and density dependences of the various operator structures in
$\vbar$.

Finally, it is possible to improve the $P=0$ approximation and to
perform the sum over the third particle self-consistently, which
corresponds to closing the line in Eq.~(\ref{vbardiag}) with the
self-consistent propagator. For finite $P$, one has more allowed spin
operators with more complex integral functions that can depend on $P$
but also on the angle of ${\bf P}$ with respect to ${\bf k}$ and ${\bf
k}'$. Since $\vbar$ has been derived using Mathematica for general
particle momenta ${\bf k}_i$, this is directly possible. One could
then explore angle-averaging over $\widehat{\bf P}$ or averaging over
the magnitude of ${\bf P}$. However, as will be shown in the next
section, the $P=0$ approximation is reliable for bulk properties and
neutron matter based on chiral low-momentum interactions is sufficiently
perturbative, which justifies using the noninteracting density to sum
over the third particle.

\section{Results}
\label{results}

We apply the developed density-dependent two-body interaction $\vbar$
to calculate the properties of neutron matter in a loop expansion
around the Hartree-Fock energy. These are the first results for
neutron matter based on chiral EFT interactions and including N$^2$LO
3N forces. The many-body calculation follows the strategy of
Refs.~\cite{nucmatt,chiralnm,neutmatt}, but with significant
improvements for the second-order contributions involving $\vbar$
and with fully self-consistent single-particle energies.

\subsection{Hartree-Fock and $P$ dependence of $\vbar$}
\label{HF}

\begin{figure}[t]
\begin{center}
\parbox[b]{2.5cm}{\centering \includegraphics[scale=0.8]{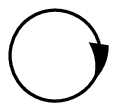}
\vspace{0.2cm} \\$E_{\rm kin}$}
\parbox[b]{2.5cm}{\centering \includegraphics[scale=0.8]{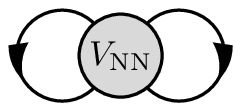}
\vspace{0.2cm} \\$E^{(1)}_{\rm NN}$}
\parbox[b]{2.5cm}{\centering \includegraphics[scale=0.8]{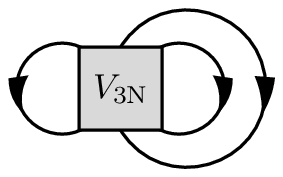}
\vspace{0.2cm} \\$E^{(1)}_{\rm 3N}$} \\[5mm]
\parbox[b]{2.5cm}{\centering \includegraphics[scale=0.8]{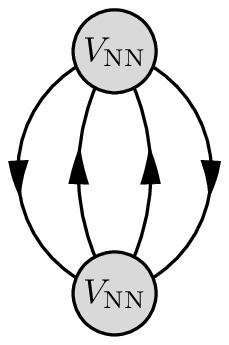}
\vspace{0.2cm} \\$E^{(2)}_{1}$}
\parbox[b]{2.5cm}{\centering \includegraphics[scale=0.8]{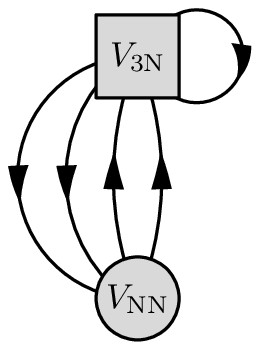}
\vspace{0.2cm} \\$E^{(2)}_{2}$}
\parbox[b]{2.5cm}{\centering \includegraphics[scale=0.8]{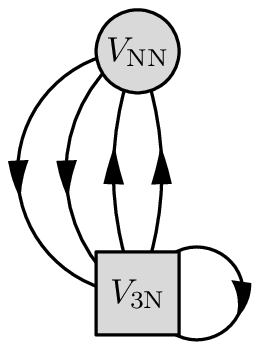}
\vspace{0.2cm} \\$E^{(2)}_{3}$} \\[5mm]
\parbox[b]{2.5cm}{\centering \includegraphics[scale=0.8]{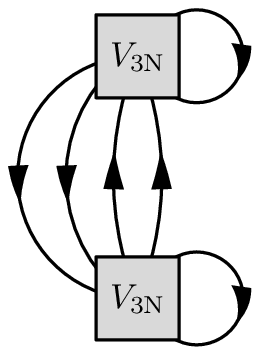}
\vspace{0.2cm} \\$E^{(2)}_{4}$}
\parbox[b]{2.5cm}{\centering \includegraphics[scale=0.8]{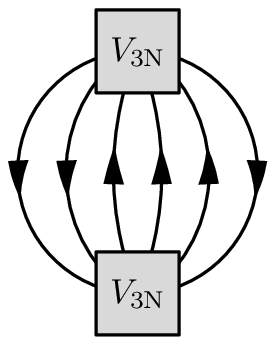}
\vspace{0.2cm} \\$E^{(2)}_{5}$}
\end{center}
\caption{Top row: Diagrams contributing to the Hartree-Fock
energy. These include the kinetic energy $E_{\rm kin}$ and the
first-order NN and 3N interaction energies, $E^{(1)}_{\rm NN}$
and $E^{(1)}_{\rm 3N}$. Middle and bottom rows: Second-order
contributions to the energy due to NN-NN interactions $E^{(2)}_{1}$,
NN-3N and 3N-3N interactions, where 3N forces enter as
density-dependent two-body interactions, $E^{(2)}_{2,3}$ and
$E^{(2)}_{4}$, respectively, and the remaining 3N-3N diagram
$E^{(2)}_{5}$.\label{fig:E_diags}}
\end{figure}

The contributions to the Hartree-Fock energy are shown
diagrammatically in Fig.~\ref{fig:E_diags} and the first-order NN and
3N interaction energies are given by
\begin{align}
\frac{E^{(1)}_{\rm NN}}{V} &= \frac{1}{2} \: \tr_{\sigma_1} \tr_{\sigma_2} 
\int \frac{d{\bf k}_1}{(2\pi)^3} \int \frac{d{\bf k}_2}{(2\pi)^3}
\nonumber \\[2mm]
&\times n_{{\bf k}_1} \, n_{{\bf k}_2} \, \langle 1 2 | \, (1-P_{12}) \,
\vlowk \bigr|_{nn} \, | 1 2 \rangle \,, \label{HFNN} \\[2mm]
\frac{E^{(1)}_{\rm 3N}}{V} &= \frac{1}{6} \: \tr_{\sigma_1}
\tr_{\sigma_2} \tr_{\sigma_3} \int \frac{d{\bf k}_1}{(2\pi)^3}
\int \frac{d{\bf k}_2}{(2\pi)^3} 
\int \frac{d{\bf k}_3}{(2\pi)^3} \nonumber \\[2mm]
&\times n_{{\bf k}_1} n_{{\bf k}_2} n_{{\bf k}_3} \, f_{\text{R}}^2(p,q) \, 
\langle 1 2 3 | \, {\mathcal A}_{123} \, V_{\rm 3N} \bigr|_{nnn}
\, | 1 2 3 \rangle \,, \label{HF3N}
\end{align}
where $V$ is the volume and we use the shorthand notation $i \equiv
{\bf k}_i \sigma_i$ in the bra and ket states. The
momentum-conserving delta functions are not included in the NN and 3N
matrix elements. It is evident from Eq.~(\ref{HF3N}) that the correct
3N symmetry factor is obtained when the antisymmetrized two-body
interaction $V^{(0)}_{\rm as} = (1-P_{12}) \, \vlowk
+ \vbar/3$ is used at the Hartree-Fock level.
With the expansion in two-body partial waves, we have
\begin{multline}
\frac{E^{(1)}_{\rm NN} + E^{(1)}_{\overline{\rm 3N}}}{V}
= \frac{1}{\pi^3} \int k^2 \, dk \int P^2 \, dP \int d\cos
\theta_{{\bf k}, {\bf P}} \\[2mm]
\times n_{\frac{\bf P}{2} + {\bf k}} \, n_{\frac{\bf P}{2} - {\bf k}}
\sum_{S,l,J} \, (2 J +1) \bigl\la k \bigl| V_{S l l J}^{(0)} \bigr| 
k \bigr\ra \bigl(1 - (-1)^{l+S+1} \bigr) \,. \label{NNpw}
\end{multline}
In Eq.~(\ref{NNpw}) and in the following, all partial-wave matrix
elements are normalized to the direct term, so that $V^{(0)}_{S l l'
J} = V_{{\rm low}\,k, S l l' J} + \overline{V}_{{\rm 3N}, S l l' J}/6$.

\begin{figure}[t]
\begin{center}
\includegraphics[scale=0.9,clip=]{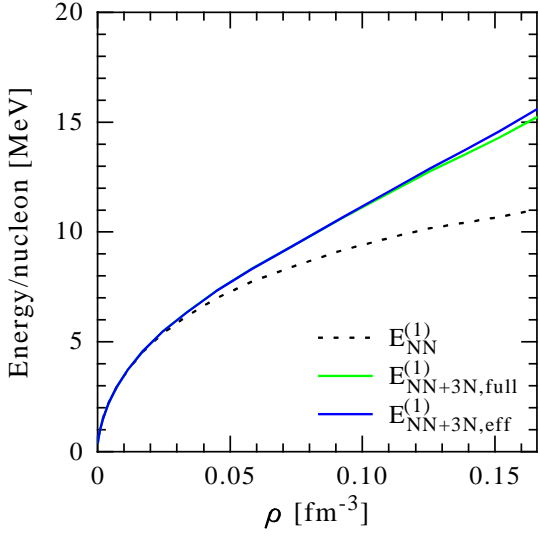}
\end{center}
\caption{(Color online) Energy per particle $E/N$ 
of neutron matter as a function of
density $\rho$ at the Hartree-Fock level, based on low-momentum 
chiral NN and 3N interactions with $\lm/\lm_{\rm 3NF} = 2.0 \fmi$. The
first-order energy $E^{(1)}_{\rm 3N,eff}$ using $\vbar$ with $P=0$
is compared to the exact Hartree-Fock
energy $E^{(1)}_{\rm 3N,full}$. To show the
impact of chiral 3N forces, we also give the first-order energy based
only on NN interactions.\label{fig:PW_test}}
\end{figure}

For N$^2$LO 3N forces, the first-order 3N interaction energy has been
calculated exactly in Ref.~\cite{neutmatt}, without the fixed-$P$
approximation in $\vbar$,
\begin{align}
&\frac{E^{(1)}_{\rm 3N}}{V} = \frac{g^2_A}{f^4_{\pi}} \: 
\int \frac{d{\bf k}_1 d{\bf k}_2 d{\bf k}_3}{(2\pi)^9} 
\: n_{{\bf k}_1} \, n_{{\bf k}_2} \, n_{{\bf k}_3} \,
f_{\text{R}}^2(p,q) \nonumber \\[2mm]
&\times \biggl[ - 2 c_1 m_\pi^2 \biggl( 
\frac{{\bf k}_{12} \cdot {\bf k}_{23}}{(k_{12}^2 + m_\pi^2)
(k_{23}^2 + m_\pi^2)} + \frac{k_{12}^2}{(k_{12}^2 + m_\pi^2)^2} \biggl)
\nonumber \\[2.5mm]
&+ c_3 \biggl( \frac{({\bf k}_{12} \cdot {\bf k}_{23})^2}{(k_{12}^2 + m_\pi^2)
(k_{23}^2 + m_\pi^2)} - \frac{k_{12}^4}{(k_{12}^2 + m_\pi^2)^2} \biggl)
\biggr] \,,
\label{3Nfull}
\end{align}
with ${\bf k}_{ij} = {\bf k}_i - {\bf k}_j$. By comparing the
Hartree-Fock energy with the density-dependent two-body interaction
$\vbar$ for $P=0$, Eq.~(\ref{NNpw}), to the result with the full
first-order 3N energy, Eq.~(\ref{3Nfull}), we can assess the
reliability of the $P=0$ approximation in $\vbar$. Our results for
$E^{(1)}_{\rm NN+3N,eff} =E_{\rm kin} + E^{(1)}_{\rm NN} +
E^{(1)}_{\overline{\rm 3N}}$ in comparison to the full Hartree-Fock
energy $E^{(1)}_{\rm NN+3N,full}$ are shown for densities $\rho < 0.16
\fmiq$ (below saturation density $\rho_0 = 0.16 \fmiq$) in
Fig.~\ref{fig:PW_test}.  The $P=0$ approximation for $\vbar$ leads to
energies that agree very well with the exact result at the
Hartree-Fock level. For saturation density ($\kf = 1.7 \fmi$), the
difference is $340 \kev$, while for lower $\kf$, the differences
are always below $100 \kev$, as documented in
Table~\ref{tab:panel_data}. This is very encouraging for evaluating
higher-order contributions, where calculations with full 3N forces
become increasingly more complex than with two-body
interactions.

As expected from the partial-wave matrix elements, we find in
Fig.~\ref{fig:PW_test} that chiral 3N forces at first order
increase the neutron matter energy with increasing density, compared
to results based only on low-momentum NN interactions. This repulsion
is due to the central parts of $\vbar$, because noncentral forces do
not contribute to the Hartree-Fock energy.

\subsection{Second order: single-particle energies}
\label{spenergies}

For the contributions beyond Hartree-Fock, we need to calculate
the single-particle energies $\varepsilon_{\bf k}$. These are
determined by the self-consistent solution to the Dyson equation.
To second order, we have
\be
\varepsilon_{\bf k} = \frac{k^2}{2m} + \Sigma^{(1)}(k)
+ {\rm Re} \, \Sigma^{(2)}(k,\varepsilon_{\bf k}) \,.
\label{eq:Dyson}
\ee
Here $\Sigma^{(1)}$ denotes the first-order NN and 3N contributions
to the self-energy, which are real, $\Sigma^{(2)}$ includes the
second-order two-particle--one-hole and two-hole--one-particle
terms, and the self-energy is averaged over the single-particle spin.
The diagrams contributing to the self-energy to second order
are given by
\be
\parbox{1.0cm}{\centering \includegraphics[scale=0.95]{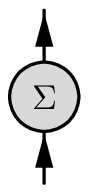}}
\parbox{0.6cm}{\centering{$=$}}
\parbox{1.4cm}{\centering \includegraphics[scale=0.95]{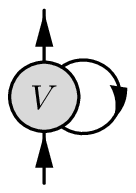}}
\parbox{0.6cm}{\centering{$+$}}
\parbox{1.4cm}{\centering \includegraphics[scale=0.95]{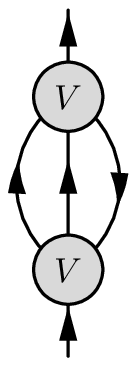}}
\ee
The resulting first- and second-order self-energies can be expressed
in terms of two-body partial waves and one has (see also Ref.~\cite{Sigma})
\begin{widetext}
\begin{align}
\Sigma^{(1)}(k_1) &= \frac{1}{2 \pi} \int k_2^2 \, dk_2
\int d\cos \theta_{{\bf k}_1,{\bf k_2}} \: n_{{\bf k}_2}
\sum_{S,l,J} \, (2 J +1) \, \bigl\la k_{12}/2 \, \bigl| V_{S l l J}^{(1)}
\bigr| \, k_{12}/2 \bigr\ra \, \bigl(1 - (-1)^{l+S+1} \bigr) \,,
\label{eq:Sigma1} \\[2mm]
\Sigma^{(2)}(k_1,\omega_1) &= \frac{4}{k_1 \pi^2} \int P \, dP 
\int k \, dk \int d{\bf k}' 
\sum_{S,M_S,M_S'} \: \sum_{l,l',l'',m,m',m''} \: \sum_{J,M} \: i^{l' - l''} \,
Y_{lm}^*\bigl(\widehat{\bf k}\bigr) \, Y_{lm}\bigl(\widehat{\bf k}\bigr)
\, Y_{l'm'}\bigl(\widehat{{\bf k}'}\bigr) \, 
Y_{l''m''}^*\bigl(\widehat{{\bf k}'}\bigr) \nonumber \\[1mm]
&\times \bigl( \mathcal{C}_{l m S M_S}^{J M} \bigr)^2 \,
\mathcal{C}_{l' m' S M'_S}^{J M} \, \mathcal{C}_{l'' m'' S M'_S}^{J M} \,
\bigl\la k \bigl| V_{S l l' J}^{(2)} \bigr| k' \bigr\ra \,
\bigl\la k' \bigl| V_{S l'' l J}^{(2)} \bigr| k \bigr\ra \,
\bigl(1 - (-1)^{l+S+1} \bigr) \, \bigl(1 - (-1)^{l'+S+1} \bigr)
\nonumber \\[2.5mm]
&\times \biggl[ \frac{(1 - n_{\frac{\bf P}{2} + {\bf k}'})
(1 - n_{\frac{\bf P}{2} - {\bf k}'}) \, n_{{\bf P} - {\bf k}_1}}{\omega_1 
+ \varepsilon_{{\bf P} - {\bf k}_1}
- \varepsilon_{\frac{\bf P}{2} + {\bf k}'} - \varepsilon_{\frac{\bf P}{2} - {\bf k}'}
+ i \delta} + \frac{n_{\frac{\bf P}{2} + {\bf k}'} \, n_{\frac{\bf P}{2} - {\bf k}'}
\, (1-n_{{\bf P} - {\bf k}_1})}{\omega_1 + \varepsilon_{{\bf P} - {\bf k}_1}
- \varepsilon_{\frac{\bf P}{2} + {\bf k}'} - \varepsilon_{\frac{\bf P}{2} - {\bf k}'}
- i \delta} \biggr] \,, \label{eq:Sigma2}
\end{align}
\end{widetext}
where $\mathcal{C}_{l m S M_S}^{J M}$ denote
Clebsch-Gordan coefficients and $Y_{lm}\bigl(\widehat{\bf k}\bigr)$
are spherical harmonics. In Eq.~(\ref{eq:Sigma2}), the $z$-axis is
taken in the ${\bf P}$ direction, we have used $k \, dk = P \, k_1
\, d\widehat{\bf k}_1/(4 \pi)$, and ${\bf k} = {\bf k}_1 - {\bf P}/2$
determines the argument $\widehat{\bf k}$.
As discussed in Sect.~\ref{operator},
the antisymmetrized two-body interactions in the first- and
second-order terms are given by $V^{(1)}_{\rm as} = (1-P_{12}) \,
\vlowk + \vbar/2$ and $V^{(2)}_{\rm as} = (1-P_{12}) \, \vlowk
+ \vbar$, with partial waves $V^{(1)}_{Sll'J}=V_{{\rm low}\,k, Sll'J}
+\overline{V}_{{\rm 3N}, Sll'J}/4$ and $V^{(2)}_{Sll'J}=
V_{{\rm low}\,k, Sll'J}+\overline{V}_{{\rm 3N}, Sll'J}/2$.

\begin{figure}[t]
\begin{center}
\includegraphics[scale=0.9,clip=]{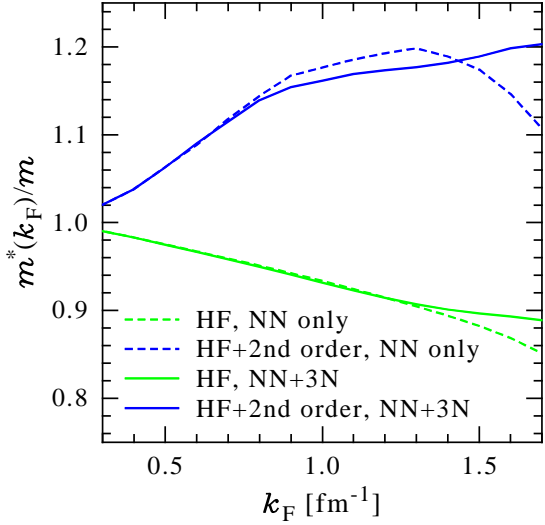}
\end{center}
\vspace*{-2mm}
\caption{(Color online) Effective mass $m^*(\kf)/m$ at the Fermi
surface as a function of Fermi momentum $\kf$ in neutron matter.
Results for $\lm/\lm_{\rm 3NF} = 2.0 \fmi$ are shown at the 
Hartree-Fock level, plus second-order contributions, and
based only on NN interactions for comparison. At second order,
the effective mass includes $k$-mass and $e$-mass effects.
\label{fig:mstar_kf}}
\end{figure}

We solve the Dyson equation, Eq.~(\ref{eq:Dyson}), self-consistently
using the self-energies given by Eqs.~(\ref{eq:Sigma1})
and~(\ref{eq:Sigma2}). In Fig.~\ref{fig:mstar_kf}, we show the
resulting effective mass at the Fermi surface,
\be
\frac{m^*(\kf)}{m} = \biggl( \frac{m}{k} \frac{d 
\varepsilon_{\bf k}}{dk} \biggr)^{-1} \biggr|_{k=\kf} \,.
\ee
At the Hartree-Fock level, 3N contributions only change the effective
mass marginally. Including second-order contributions leads to the
typical enhancement of the effective mass at the Fermi surface, and we
find a larger impact of 3N forces for $\kf > 1.3 \fmi$.

\subsection{Second order: energy per particle}
\label{2nd}

\begin{figure*}[t]
\begin{center}
\includegraphics[scale=0.9,clip=]{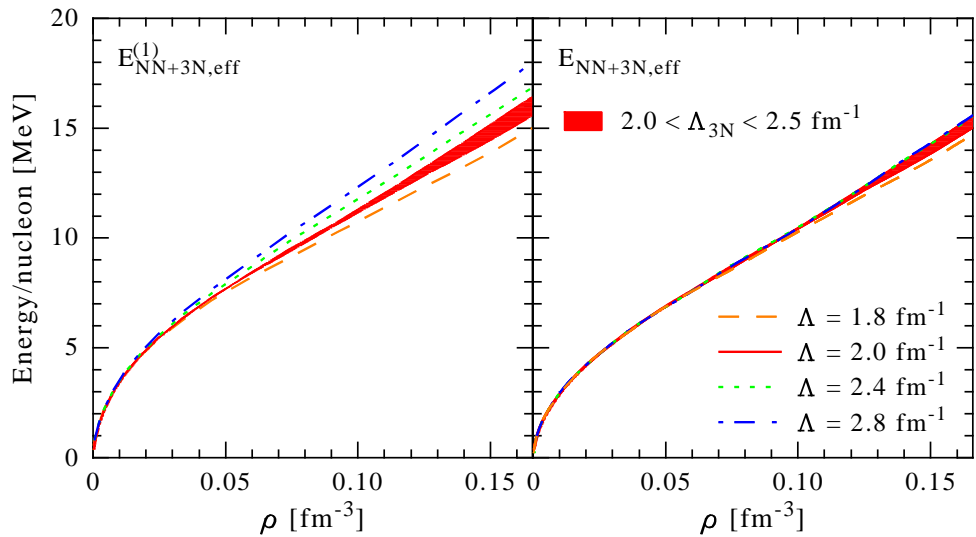}
\end{center}
\caption{(Color online) Energy per particle $E/N$ of neutron matter
as a function of density $\rho$ at the Hartree-Fock level (left) and
including second-order contributions (right). The results are based
on evolved N$^3$LO NN potentials and N$^2$LO 3N forces. Theoretical
uncertainties are estimated by varying the NN cutoff (lines) and the
3N cutoff (band for fixed $\lm = 2.0 \fmi$).\label{fig:panel}}
\end{figure*}

We include the second-order contributions $E^{(2)}_1$ to
$E^{(2)}_4$ of Fig.~\ref{fig:E_diags}, which are given by
\begin{multline}
E^{(2)}_{\rm NN+3N,eff} = \frac{1}{4} \, \biggl( \, \prod_{i=1}^{4} \,
\tr_{\sigma_i} \int \frac{d{\bf k}_i}{(2\pi)^3} \biggr) \,
\bigl| \langle 1 2 \, \bigl| \, V^{(2)}_{\rm as} \, \bigr| \,
3 4 \rangle \bigr|^2 \\[1mm]
\times \frac{n_{{\bf k}_1} n_{{\bf k}_2}
(1-n_{{\bf k}_3}) (1-n_{{\bf k}_4})}{
\varepsilon_{{\bf k}_1} + \varepsilon_{{\bf k}_2}
- \varepsilon_{{\bf k}_3} - \varepsilon_{{\bf k}_4}}
\, (2 \pi)^3 \, \delta({\bf k}_1 + {\bf k}_2 - {\bf k}_3 - {\bf k}_4)
\,. \label{E2}
\end{multline}
As in the second-order self-energy, the antisymmetrized two-body
interactions when evaluating contributions beyond the Hartree-Fock
level are given by $V^{(2)}_{\rm as} = (1-P_{12}) \, \vlowk + \vbar$.
The second-order calculations are carried out using the
self-consistent single-particle energies determined by solving the
Dyson equation, Eq.~(\ref{eq:Dyson}), as discussed in
Sect.~\ref{spenergies}, and the intermediate-state phase-space
integrations are performed fully. Summing over the spins and
expanding in partial waves, we have~\cite{neutmatt}
\begin{align}
&\sum_{S, M_S, M'_S} \bigl|
\langle {\bf k} \, S M_S \, | \, V^{(2)}_{\rm as} \, | \, {\bf k'} \,
S M'_S \rangle \bigr|^2 = \sum_L \, P_L(\cos \theta_{{\bf k},{\bf k'}})
\nonumber \\[1mm]
&\times \sum_{J,l,l',S} \, \sum_{\tj,\tl,\tlp} \, (4 \pi)^2 \, 
i^{(l-l'+\tl-\tlp)} \, (-1)^{\tl+l'+L} \, 
\mathcal{C}_{l 0 \tlp 0}^{L 0} \, \mathcal{C}_{l' 0 \tl 0}^{L 0}
\nonumber \\[1mm]
&\times \sqrt{(2 l + 1) (2 l' + 1) (2 \tl + 1) (2 \tlp + 1)} 
\: (2J+1) (2\tj+1) \nonumber \\[2mm]
&\times \biggl\{ \begin{array}{ccc}
l & S & J \\
\tj & L & \tlp \end{array} \biggr\}
\biggl\{ \begin{array}{ccc}
J & S & l' \\
\tl & L & \tj \end{array} \biggr\} \,
\bigl\la k \bigl| V_{S l' l J}^{(2)} \bigr| k' \bigr\ra \,
\bigl\la k' \bigl| V_{S \tlp \tl \tj}^{(2)} \bigr| k \bigr\ra
\nonumber \\[1.5mm]
& \times \bigl(1-(-1)^{l+S+1}\bigr) \, \bigl(1-(-1)^{\tl+S+1}\bigr) \,,
\label{recoupling}
\end{align}
where $\{ \ldots \}$ denote 6j-symbols and $P_L(\cos \theta)$ are
Legendre polynomials. Keeping only $L=0$ in Eq.~(\ref{recoupling})
corresponds to the angle-averaging approximation for the
Pauli-blocking operator, but we keep all $L \leqslant 6$ for $\vlowk$
and $L \leqslant 4$ for $\vbar$.

\begin{table*}[t]
\begin{tabular}{c|c||c|c|c|c|c|c|c}
\hline\hline
\hspace*{1mm} $\kf$ \hspace*{1mm} & 
\hspace*{1mm} $\lm/\lm_{\rm 3NF}$ \hspace*{1mm} & 
\hspace*{1mm} $E_{\rm kin}$ \hspace*{1mm} & 
\hspace*{1mm} $E^{(1)}_{\rm NN}$ \hspace*{1mm} & 
\hspace*{1mm} $E^{(1)}_{\rm 3N,full}$ \hspace*{1mm} & 
\hspace*{1mm} $E^{(1)}_{\rm 3N,eff}$ \hspace*{1mm} & 
\hspace*{1mm} $E^{(2)}_1$ \hspace*{1mm} & 
\hspace*{1mm} $E^{(2)}_2 + E^{(2)}_3$ \hspace*{1mm} & 
\hspace*{1mm} $E^{(2)}_4$ \hspace*{1mm} \\
\hline
1.3 & 1.8/2.0 & 21.01 & -12.86 & 0.95 & 0.94 & -0.59 & 0.01 & -0.02 \\
1.3 & 2.0/2.0 & 21.01 & -12.58 & 0.95 & 0.94 & -0.78 & 0.00 & -0.02 \\
1.3 & 2.0/2.5 & 21.01 & -12.58 & 1.05 & 1.00 & -0.77 & -0.01 & -0.05 \\
1.3 & 2.4/2.0 & 21.01 & -12.11 & 0.95 & 0.94 & -1.10 & -0.02 & -0.02 \\
1.3 & 2.8/2.0 & 21.01 & -11.75 & 0.95 & 0.94 & -1.46 & -0.03 & -0.02 \\
\hline
1.5 & 1.8/2.0 & 27.97 & -18.62 & 2.18 & 2.24 & -0.39 & 0.01 & -0.05 \\
1.5 & 2.0/2.0 & 27.97 & -18.14 & 2.18 & 2.24 & -0.64 & -0.01 & -0.05 \\
1.5 & 2.0/2.5 & 27.97 & -18.14 & 2.56 & 2.51 & -0.63 & -0.04 & -0.14 \\
1.5 & 2.4/2.0 & 27.97 & -17.44 & 2.18 & 2.24 & -1.16 & -0.05 & -0.05 \\
1.5 & 2.8/2.0 & 27.97 & -16.77 & 2.18 & 2.24 & -1.78 & -0.08 & -0.05 \\
\hline
1.7 & 1.8/2.0 & 35.93 & -25.50 & 4.20 & 4.54 & -0.22 & 0.01 & -0.07 \\
1.7 & 2.0/2.0 & 35.93 & -24.93 & 4.20 & 4.54 & -0.45 & -0.02 & -0.08 \\
1.7 & 2.0/2.5 & 35.93 & -24.93 & 5.36 & 5.40 & -0.46 & -0.06 & -0.31 \\
1.7 & 2.4/2.0 & 35.93 & -23.64 & 4.20 & 4.54 & -1.11 & -0.07 & -0.08 \\
1.7 & 2.8/2.0 & 35.93 & -22.51 & 4.20 & 4.54 & -2.08 & -0.12 & -0.09 \\
\hline\hline
\end{tabular}
\caption{Contributions to the neutron matter energy due to the
diagrams of Fig.~\ref{fig:E_diags}. Results are given for Fermi
momenta $\kf = 1.3, 1.5$ and $1.7 \fmi$ and for different
$\lm/\lm_{\rm 3NF}$ combinations. All energies are in ${\rm MeV}$
and $\kf$, $\lm/\lm_{\rm 3NF}$ are in ${\rm fm}^{-1}$.
\label{tab:panel_data}}
\end{table*}

Our second-order results for the neutron matter energy, $E_{\rm
  NN+3N,eff} = E^{(1)}_{\rm NN+3N,eff} + E^{(2)}_{\rm NN+3N,eff}$, are
presented in Fig.~\ref{fig:panel}. The different contributions are
listed in Table~\ref{tab:panel_data}. We observe that the cutoff
dependence is reduced when going from first to second order. This is
as expected based on the nuclear matter
results~\cite{nucmatt,chiralnm}, but for neutron matter the cutoff
dependence is significantly weaker already at the Hartree-Fock level.
The cutoff dependence increases with density and is less than $1
\mev$ per particle for the densities studied in Fig.~\ref{fig:panel}
over the cutoff range $1.8 \fmi \leqslant \lm \leqslant 2.8 \fmi$ and
$2.0 \fmi \leqslant \Lambda_{\rm 3NF} \leqslant 2.5 \fmi$. This band sets
the scale for omitted short-range many-body contributions and we
discuss the theoretical uncertainties in the long-range parts in the
following section. The weak cutoff dependence also demonstrates
that the average momentum in the system (which is smaller than the
Fermi momentum, because $\langle p^2_i \rangle = 3/5 \, \kf^2$) is
well below the cutoff.

Moreover, we have found that self-energy corrections to the neutron
matter energy are practically negligible. The second-order energy with
$\Sigma = 0$ is within $200 \kev$ of the self-consistent results shown
in Fig.~\ref{fig:panel}. In addition, the second-order energy
contributions are always below $1.3 \mev$ per particle in
Table~\ref{tab:panel_data}, except for the large cutoff $\lm = 2.8
\fmi$ cases. The second-order contributions
practically only improve the cutoff independence of the results
without changing the energy significantly. Moreover, for the lower
cutoffs, the Hartree-Fock energies are already reliable. These
findings combined suggest that neutron matter is perturbative at
nuclear densities. Therefore, we are confident that the $P=0$
approximation for $\vbar$ is reliable, when evaluating the small
second-order contributions, and that it is reasonable to neglect the
residual 3N-3N diagram, $E^{(2)}_5$.

\subsection{Sensitivity to $c_i$ uncertainties}
\label{ciuncertain}

Next, we study the sensitivity of the second-order energy to
uncertainties in the $c_i$ coefficients that determine the long-range
part of N$^2$LO 3N forces. This provides an update for chiral
potentials of the results of Ref.~\cite{neutmatt}. The $c_i$
coefficients relate $\pi$N, NN and 3N interactions, and the
determination from $\pi$N scattering is, within errors, consistent
with the extraction from NN waves. Present constraints for $c_1$ and
$c_3$ are $c_1 = -0.9^{+0.2}_{-0.5} \, {\rm GeV}^{-1}$ and $c_3 =
-4.7^{+1.5}_{-1.0} \, {\rm GeV}^{-1}$~\cite{Ulf}. We note that
at N$^3$LO there are contributions that shift the $c_i$~\cite{RMP},
and may lead to $c_3$ coefficients that are smaller in magnitude. In
this study, we vary $c_i$ only in 3N forces, because of lack of
N$^3$LO NN potentials that explore these $c_i$ variations. However,
based on the universality of $\vlowk$~\cite{chiralnm,smooth} (starting
from chiral potentials with two different $c_i$
sets~\cite{N3LO,N3LOEGM}), we do not expect large differences from
varying $c_1$ and $c_3$ in NN interactions, where these variations are
also absorbed by higher-order contact interactions that have to be
adjusted to reproduce NN scattering.

\begin{figure}[t]
\begin{center}
\includegraphics[scale=0.9,clip=]{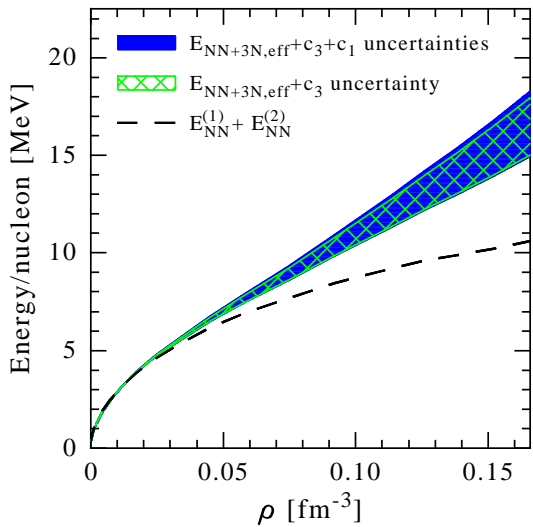}
\end{center}
\caption{(Color online) Theoretical uncertainties of the second-order 
energy with $\lm/\lm_{\rm 3NF} = 2.0 \fmi$
as a function of density due to the uncertainties in the $c_1$
and $c_3$ coefficients of 3N forces.\label{fig:ci_uncert}}
\end{figure}

In Fig.~\ref{fig:ci_uncert} we show that the theoretical uncertainties
of the neutron matter energy are dominated by the uncertainties in the
$c_i$ coefficients, in particular the $c_3$ part, compared to the
uncertainties of the many-body calculation or of neglected
short-range many-body interactions probed by cutoff variations. The
$c_1$ and $c_3$ variation leads to an energy uncertainty of $\pm 1.5
\mev$ per particle at saturation density. Therefore, it is important
to improve the constraints on $c_i$. Figure~\ref{fig:ci_uncert} also
shows that N$^2$LO 3N forces provide a repulsive contribution to
neutron matter at second order, even with this band.

\begin{figure}[t]
\begin{center}
\includegraphics[scale=0.9,clip=]{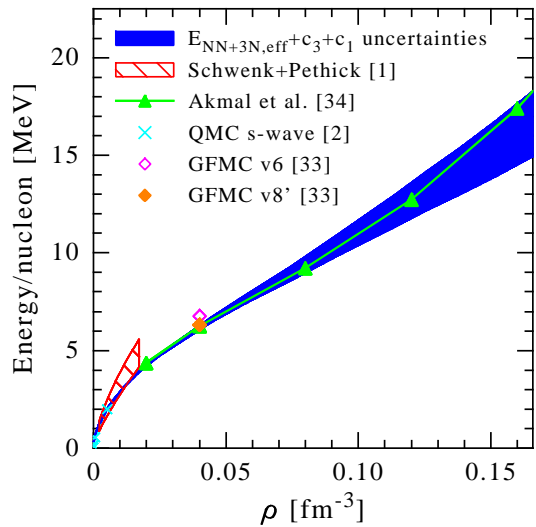}
\end{center}
\caption{(Color online) Comparison of the second-order energy
with the $c_i$ uncertainty band of Fig.~\ref{fig:ci_uncert}
to other neutron matter results (see text for details).
\label{fig:EOS_compare}}
\end{figure}

We compare our neutron matter energy with the $c_i$ uncertainty band to
other recent neutron matter results in Fig.~\ref{fig:EOS_compare}.
These include Monte-Carlo calculations (GFMC $v_6$ and
$v_8'$~\cite{Carlson} and QMC $S$-wave~\cite{Gezerlis}) and di-fermion
EFT results~\cite{dEFT} for lower densities, which are only sensitive
to parts of NN interactions, at very low densities only to the
neutron-neutron scattering length and effective range. In addition, we
include in Fig.~\ref{fig:EOS_compare} the results of Akmal {\it et
al.}~\cite{Akmal} based on the Argonne $v_{18}$ NN and Urbana IX 3N
potentials, where the repulsive 3N contributions arise from various terms.

\subsection{Symmetry energy}
\label{symenergy}

The spread in the energy band of Fig.~\ref{fig:ci_uncert} corresponds
to a range for the symmetry energy and its density dependence. In the
following we explore this correlation and consequently also a
possibility of using nuclear matter properties to constrain some of
the $c_i$ couplings. As a function of asymmetry $\alpha =
(\rho_n-\rho_p)/(\rho_n+\rho_p)$, the energy of nuclear matter can be
parameterized around saturation density as (see for example,
Refs.~\cite{sym1,sym2})
\be
\frac{E(\rho,\alpha)}{A} = - a_V +\frac{K}{18 \rho_0^2} \, (\rho - \rho_0)^2 
+ S_2 (\rho) \, \alpha^2 + \ldots \,,
\label{Eform}
\ee
with the density-dependent symmetry energy $S_2(\rho)$,
\be
S_2(\rho) = a_4 + \frac{p_0}{\rho_0^2} \, (\rho - \rho_0) + \ldots \,.
\ee
Here $A$ is the nucleon number, $a_V$ the binding energy of symmetric
nuclear matter at saturation density, $K$ is the incompressibility,
$a_4$ the symmetry energy and $p_0$ the linear dependence of the
symmetry energy.

\begin{table}[t]
\begin{tabular}{c|c||c|c}
\hline\hline
\hspace*{0.5mm} $c_1$ [${\rm GeV}$] \hspace*{0.5mm} &
\hspace*{0.5mm} $c_3$ [${\rm GeV}$] \hspace*{0.5mm} &
\hspace*{0.5mm} $a_4$ [${\rm MeV}$] \hspace*{0.5mm} &
\hspace*{0.5mm} $p_0$ [${\rm MeV \, fm}^{-3}$] \hspace*{0.5mm} \\
\hline
$-0.81$ & $-3.2$ & $30.5$ & $2.0/2.0$ \\
$-0.81$ & $-5.7$ & $33.2$ & $2.8/2.8$ \\
\hline
$-0.7$ & $-3.2$ & $30.4$ & $2.0/2.0$ \\
$-1.4$ & $-5.7$ & $33.6$ & $2.8/2.9$ \\
\hline\hline
\end{tabular}
\caption{Symmetry energy $a_4$ and the linear dependence of the symmetry
energy $p_0$ obtained from the energy band of Fig.~\ref{fig:ci_uncert}
for different $c_1$ and $c_3$ coefficients using $a_V = 16 \mev$ and
$K = 190/240 \mev$.\label{tab:a4}}
\end{table}

Using the band of Fig.~\ref{fig:ci_uncert} over the density range
$0.13 \fmiq < \rho < 0.19 \fmiq$ with empirical values for the
saturation point $\rho_0 = 0.16 \fmiq$, $a_V = 16 \mev$ and
incompressibilities $K = 190/240 \mev$ (which are in the range of
empirical values~\cite{sym2} and were obtained for the same
interactions in Ref.~\cite{chiralnm}), we find in Table~\ref{tab:a4}
that the symmetry energy ranges from $a_4 = (30.4 - 33.6) \mev$
and the linear dependence of the symmetry energy from $p_0 = (2.0 
- 2.9) \, {\rm MeV \, fm}^{-3}$. The value for
$p_0$ is practically independent of the values of $K$. As shown in
Table~\ref{tab:a4}, the resulting range for the symmetry energy and
its density dependence is set by the uncertainty of the dominant
$c_3$. Compared to the empirical range $a_4 = (25 - 35)
\mev$~\cite{sym2}, the microscopic range of $\sim 3 \mev$ is very
useful and the comparison also suggests that smaller values in
magnitude for $c_3$ may be somewhat favored.

\subsection{$^1$S$_0$ pairing gap}
\label{pairing}

Three-nucleon forces also impact superfluid pairing gaps or the
anomalous self-energy. Here we will focus on the change of the
neutron-neutron $^1$S$_0$ pairing gap $\Delta$ due to chiral 3N forces
compared to the gap based on NN interactions. For simplicity, we
neglect induced interactions and study the pairing gap at
the BCS level. In this approximation, the contribution of 3N forces to
the pairing interaction originates from two paired particles on the
Fermi surface in back-to-back momentum states and the third particle
is summed over occupied states in the superfluid BCS ground state. This
means that the gap equation is given diagrammatically by
\be
\parbox{1.0cm}{\centering \includegraphics{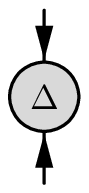}}
\parbox{0.6cm}{\centering{$=$}}
\parbox{1.4cm}{\centering \includegraphics{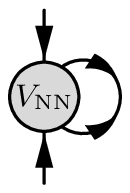}}
\parbox{0.6cm}{\centering{$+$}}
\parbox{2.0cm}{\centering \includegraphics{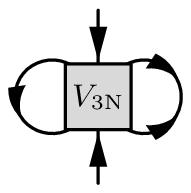}}
\label{gapdiag}
\ee
The kinematics in $V_{\rm 3N}$ is always $P=0$ for the interacting
particle pair in the gap equation and therefore taking $P=0$
in $\vbar$ is exact in this case. The only approximation using $\vbar$
at the BCS level consists in having performed the sum over occupied
states in the normal Fermi sea. However, the difference to summing
over occupied states in the BCS state is of higher order
in $\Delta/\varepsilon_{\rm F}$ ($\varepsilon_{\rm F} = \kf^2/(2m)$
being the Fermi energy), which is small in neutron matter especially
at the densities where 3N forces are effective.

In the BCS approximation, the $^1$S$_0$ superfluid gap $\Delta(k)$ is
obtained by solving the gap equation,
\begin{equation}
\Delta(k) = - \frac{1}{\pi} \int dp \, p^2 \:
\frac{V^{(2)}_{\text{$^1$S$_0$}}(k,p) \,
\Delta(p)}{\sqrt{(\varepsilon_{\bf p} - \mu)^2 + \Delta^2(p)}} \,,
\label{eq:gap_3N}
\end{equation}
where the chemical potential $\mu = \varepsilon_{\kf}$ at the BCS
level and the pairing interaction is given by the two-body
interaction including $\vbar$ in the $^1$S$_0$ channel, so that
$V^{(2)}_{\text{$^1$S$_0$}} = V_{{\rm low}\,k, \text{$^1$S$_0$}} +
\overline{V}_{{\rm 3N}, \text{$^1$S$_0$}}/2$. The gap equation
including 3N forces, Eq.~(\ref{eq:gap_3N}), can be derived
explicitly by minimizing the expectation value of the Hamiltonian
in the BCS state~\cite{Yang}.

\begin{figure}[t]
\begin{center}
\includegraphics[scale=0.9,clip=]{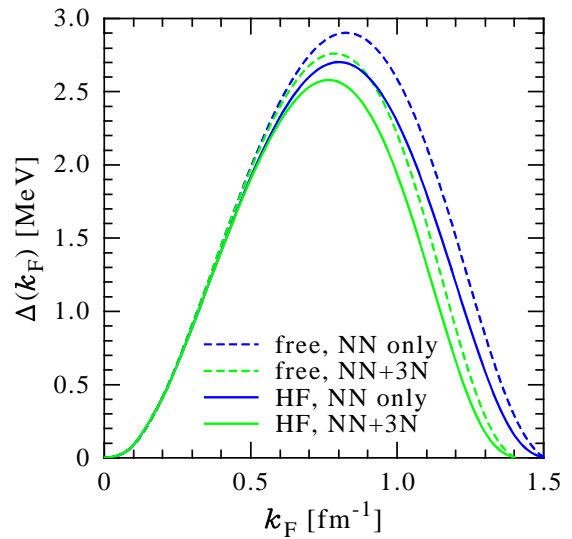}
\end{center}
\caption{(Color online) The neutron-neutron $^1$S$_0$ superfluid
pairing gap on the Fermi surface $\Delta(\kf)$ as a funtion of
Fermi momentum $\kf$. The results are based on evolved N$^3$LO
NN potentials and N$^2$LO 3N forces with $\lm/\lm_{\rm 3NF} = 2.0
\fmi$ using free and Hartree-Fock single-particle energies.
\label{fig:1S0}}
\end{figure}

We consider two cases for the single-particle energies, a free
spectrum $\varepsilon_{\bf k} = k^2/(2m)$ and Hartree-Fock
single-particle energies $\varepsilon^{(1)}_{\bf k} = k^2/(2m) +
\Sigma^{(1)}(k)$, and solve the gap equation, Eq.~(\ref{eq:gap_3N}),
as in Ref.~\cite{Kai}. Our results for the neutron-neutron pairing gap
on the Fermi surface $\Delta(\kf)$ are presented in Fig.~\ref{fig:1S0}
as a function of the Fermi momentum.  The inclusion of N$^2$LO 3N
forces leads to a reduction of the pairing gap for Fermi momenta $\kf
> 0.6 \fmi$, as expected since the $^1$S$_0$ channel of $\vbar$ is
repulsive in Fig.~\ref{fig:vbar1S0}.  This reduction agrees
qualitatively with results based on 3N potential models, see for
example, Ref.~\cite{Zuo}. While the impact of 3N forces seems small in
Fig.~\ref{fig:1S0}, for the densities, where the gap is decreasing,
the reduction due to 3N forces is significant. Figure~\ref{fig:1S0}
also shows that the effects of 3N forces relative to NN interactions is
very similar for free and Hartree-Fock single-particle energies.

\section{Conclusions and outlook}
\label{concl}

In summary, we have shown that only the $c_1$ and $c_3$ terms of the
long-range $2\pi$-exchange part of the leading chiral 3N forces
contribute in neutron matter. Based on these parts and including all
exchange terms, we derived density-dependent two-body interactions
$\vbar$ for general momentum and spin configurations by summing the
third particle over occupied states in the Fermi sea. The resulting
$\vbar$ was found to be dominated by the repulsive central part. The
comparison to the exact Hartree-Fock energy demonstrated that the $P$
dependence is very weak and we have therefore taken $P=0$ in
$\vbar$. In addition, the partial-wave matrix elements of our full
$\vbar$ approximately agree with the density-dependent two-body forces
derived recently in the framework of in-medium chiral perturbation
theory with certain approximations in Ref.~\cite{Holt}.

The density-dependent two-body interactions $\vbar$ correspond to the
normal-ordered two-body part of 3N forces and we stressed that there
are different normal ordering or symmetry factors when $\vbar$ is used
at the Hartree-Fock, one- or two-body level. Moreover, it is important
that the summation of the third particle over occupied states is
performed in the basis consistent with the one used in the many-body
calculation.

We have presented the first results for neutron matter based on chiral
EFT interactions and including N$^2$LO 3N forces. The RG evolution of
N$^3$LO NN potentials to low momenta rendered the many-body
calculation more controlled and our results for the energy suggest
that neutron matter is perturbative at nuclear densities. This is
based on small second-order contributions, self-energy corrections
being negligible, and a generally weak cutoff dependence. We have
found that N$^2$LO 3N forces provide a repulsive contribution to the
energy due to the repulsive central part in $\vbar$.  An important
direction for future work is understanding the connection to the
repulsive 3N contributions to the oxygen isotopes discovered in
Ref.~\cite{oxygen}.

We have studied in detail the theoretical uncertainties of the neutron
matter energy and found that the uncertainty in the $c_3$ coefficient
of 3N forces dominates compared to other many-body uncertainties. This
resulted in an energy band in Fig.~\ref{fig:EOS_compare}. Other
recent neutron matter calculations were found to lie within this band.
Moreover, the energy band provides microscopic constraints for the
symmetry energy and its density dependence, and thus for the neutron
skin in $^{208}$Pb~\cite{skin}. Finally, we have obtained
a significant reduction of the $^1$S$_0$ superfluid pairing gap due to
3N forces for densities where the gap is decreasing. Our results show
that chiral EFT and RG interactions can provide useful constraints for
developing a universal density functional based on microscopic
interactions~\cite{DFT1,DFT2,Wolfram}.

\acknowledgments

We thank S.\ K.\ Bogner and R.\ J.\ Furnstahl for useful discussions
and are grateful to J.\ W.\ Holt for the helpful comparison of $\vbar$
partial-wave matrix elements. This work was supported in part by the
Natural Sciences and Engineering Research Council of Canada (NSERC)
and by the Helmholtz Alliance Program of the Helmholtz Association,
contract HA216/EMMI ``Extremes of Density and Temperature: Cosmic
Matter in the Laboratory''. TRIUMF receives funding via a contribution
through the National Research Council Canada.


\begin{thebibliography}{99}
\bibitem{dEFT} A.\ Schwenk and C.\ J.\ Pethick, Phys.\ Rev.\ Lett.\ 
{\bf 95}, 160401 (2005).

\bibitem{Gezerlis} A.\ Gezerlis and J.\ Carlson, \prc {\bf 77},
032801 (2008).

\bibitem{coldatoms} S.\ Giorgini, L.\ P.\ Pitaevskii and S.\ Stringari,
\rmp {\bf 80}, 1215 (2008).

\bibitem{DFT1} S.\ K.\ Bogner, R.\ J.\ Furnstahl and L.\ Platter,
Eur.\ Phys.\ J.\ A {\bf 39}, 219 (2009).

\bibitem{DFT2} M.\ Stoitsov, J.\ More, W.\ Nazarewicz, J.\ C.\ Pei,
J.\ Sarich, N.\ Schunck, A.\ Staszczak and S.\ Wild, J.\ of
Physics: Conf.\ Series {\bf 180}, 012082 (2009).

\bibitem{LP} J.\ M.\ Lattimer and M.\ Prakash, Astrophys.\ J.\
{\bf 550} (2001) 426.

\bibitem{nucmatt} S.\ K.\ Bogner, A.\ Schwenk, R.\ J.\ Furnstahl and A.\
Nogga, Nucl.\ Phys.\ A {\bf 763}, 59 (2005).

\bibitem{chiralnm} S.\ K.\ Bogner, R.\ J.\ Furnstahl, A.\ Nogga and
A.\ Schwenk, arXiv:0903.3366 [nucl-th].

\bibitem{chiral} E.\ Epelbaum, Prog.\ Part.\ Nucl.\ Phys.\ {\bf 57},
654 (2006).

\bibitem{RMP} E.\ Epelbaum, H.-W.\ Hammer and U.-G.\ Mei{\ss}ner,
\rmp {\bf 81}, 1773 (2009).

\bibitem{Vlowk} S.\ K.\ Bogner, T.\ T.\ S.\ Kuo and A.\ Schwenk, Phys.\
Rept.\ {\bf 386}, 1 (2003); S.\ K.\ Bogner, A.\ Schwenk, T.\ T.\ S.\
Kuo and G.\ E.\ Brown, nucl-th/0111042.

\bibitem{smooth} S.\ K.\ Bogner, R.\ J.\ Furnstahl, S.\ Ramanan and A.\
Schwenk, Nucl.\ Phys.\ A {\bf 784}, 79 (2007).

\bibitem{NCSM} S.\ K.\ Bogner, R.\ J.\ Furnstahl, P.\ Maris, R.\ J.\
Perry, A.\ Schwenk and J.\ P.\ Vary, Nucl.\ Phys.\ A {\bf 801}, 21 (2008).

\bibitem{Sonia} S.\ Bacca, A.\ Schwenk, G.\ Hagen and T.\ Papenbrock,
Eur.\ Phys.\ J.\ A {\bf 42}, 553 (2009).

\bibitem{N3LO} D.\ R.\ Entem and R.\ Machleidt, Phys.\ Rev.\ C {\bf 68}, 
041001(R) (2003).

\bibitem{N3LOEGM} E.\ Epelbaum, W.\ Gl\"ockle and U.-G.\ Mei{\ss}ner,
Nucl.\ Phys.\ A {\bf 747}, 362 (2005).

\bibitem{chiral3N1} U.\ van Kolck, Phys.\ Rev.\ C {\bf 49}, 2932 (1994).

\bibitem{chiral3N2} E.\ Epelbaum, A.\ Nogga, W.\ Gl\"ockle, H.\ Kamada,
U.-G.\ Mei{\ss}ner and H.\ Wita{\l}a, Phys.\ Rev.\ C {\bf 66}, 064001
(2002).

\bibitem{3Nold1} D.\ W.\ Batt and B.\ H.\ J.\ McKellar, \prc {\bf 11},
614 (1974).

\bibitem{3Nold2} S.\ A.\ Coon, N.\ D.\ Scadron, P.\ C.\ McNamee and
B.\ R.\ Barrett, Nucl.\ Phys.\ A {\bf 317}, 242 (1979).

\bibitem{pionless} P.\ F.\ Bedaque and U.\ van Kolck, Annu.\ Rev.\ Nucl.\ 
Part.\ Sci.\ {\bf 52}, 339 (2002).

\bibitem{Kai} K.\ Hebeler, A.\ Schwenk and B.\ Friman, Phys.\ Lett.\ B
{\bf 648}, 176 (2007).

\bibitem{Born} S.\ K.\ Bogner, R.\ J.\ Furnstahl, S.\ Ramanan and A.\
Schwenk, Nucl.\ Phys.\ A {\bf 773}, 203 (2006).

\bibitem{Vlowk3N} A.\ Nogga, S.\ K.\ Bogner and A.\ Schwenk, Phys.\ Rev.\
C {\bf 70}, 061002(R) (2004).

\bibitem{neutmatt} L.\ Tolos, B.\ Friman and A.\ Schwenk, Nucl.\
Phys.\ A {\bf 806}, 105 (2008).

\bibitem{Negele} J.\ W.\ Negele and H.\ Orland, {\it Quantum 
Many-Particle Systems} (Advanced Book Classics, Westview Press, 1998).

\bibitem{nord} For normal ordering with 3N forces, see Eq.~(2) in 
G.\ Hagen, T.\ Papenbrock, D.\ J.\ Dean, A.\ Schwenk, A.\ Nogga, 
M.\ W{\l}och and P.\ Piecuch, Phys.\ Rev.\ C {\bf 76}, 034302 (2007);
or Sect.~4.3 in S.\ K.\ Bogner, R.\ J.\ Furnstahl and A.\ Schwenk,
arXiv:0912.3688 [nucl-th].

\bibitem{lit1} M.\ Baldo, I.\ Bombaci and G.\ F.\ Burgio, Astron.\ 
Astrophys.\ {\bf 328}, 274 (1997).

\bibitem{lit2} W.\ Zuo, A.\ Lejeune, U.\ Lombardo and J.\ F.\ Mathiot,
Nucl.\ Phys.\  A {\bf 706}, 418 (2002).

\bibitem{lit3} V.\ Soma and P.\ Bozek, \prc {\bf 78}, 054003 (2008).

\bibitem{Sigma}	W.\ Zuo, I.\ Bombaci and U.\ Lombardo, \prc {\bf 60},
024605 (1999). 

\bibitem{Ulf} U.-G.\ Mei{\ss}ner, private communication (2007).
We have enlarged the upper limit for $c_3$ to include the value used in
Ref.~\cite{N3LO}.

\bibitem{Carlson} J.\ Carlson, J.\ Morales, V.\ R.\ Pandharipande and
D.\ G.\ Ravenhall, \prc {\bf 68}, 025802 (2003).

\bibitem{Akmal} A.\ Akmal, V.\ R.\ Pandharipande and D.\ G.\ 
Ravenhall, Phys.\ Rev.\  C {\bf 58}, 1804 (1998).

\bibitem{sym1} R.\ J.\ Furnstahl, Nucl.\ Phys.\ A {\bf 706}, 85 (2002).

\bibitem{sym2} A.\ W.\ Steiner, M.\ Prakash, J.\ M.\ Lattimer and
P.\ J.\ Ellis, Phys.\ Rept.\ {\bf 411}, 325 (2005).

\bibitem{Yang} L.\ Yuan, Ph.D. thesis (Washington University,
St. Louis, 2006).

\bibitem{Zuo} W.\ Zuo, Z.\ H.\ Li, G.\ C.\ Lu, J.\ Q.\ Li, W.\
Scheid, U.\ Lombardo, H.-J.\ Schulze and C.\ W.\ Shen, Phys.\ Lett.\
B {\bf 595}, 44 (2004).

\bibitem{Holt} J.\ W.\ Holt, N.\ Kaiser and W.\ Weise, arXiv:0910.1249
[nucl-th].

\bibitem{oxygen} T.\ Otsuka, T.\ Suzuki, J.\ D.\ Holt, A.\ Schwenk and
Y.\ Akaishi, arXiv:0908.2607 [nucl-th].

\bibitem{Wolfram} N.\ Kaiser, S.\ Fritsch and W.\ Weise, Nucl.\ Phys.\
A {\bf 724}, 47 (2003).

\bibitem{skin} K.\ Hebeler and A.\ Schwenk, in preparation.
\end{thebibliography}
\end{document}